\documentclass[pra,twocolumn,superscriptaddress,showpacs,amsmath,amstex,amssymb,citeautoscript]{revtex4-1}

\usepackage[T1]{fontenc}
\usepackage[utf8]{inputenc}
\usepackage{lipsum}
\usepackage{amsmath}
\usepackage{amssymb}
\usepackage{bbm}
\usepackage{braket}
\usepackage{xcolor}
\usepackage{pifont}
\usepackage{footmisc}
\newcommand{\cmark}{\ding{51}}%
\newcommand{\xmark}{\ding{55}}%
\usepackage[mathscr]{euscript}
\usepackage[shortlabels]{enumitem}
\usepackage[justification=justified]{subcaption}
\usepackage{graphicx}
\usepackage{lipsum}
\allowdisplaybreaks
\usepackage{float}
\usepackage{graphicx}
\usepackage{dsfont}
\usepackage{comment}
\usepackage[colorlinks=true]{hyperref}  
\hypersetup{
    bookmarks=true,         
    unicode=false,          
    pdftoolbar=true,        
    pdfmenubar=true,        
    pdffitwindow=false,     
    pdfstartview={FitH},    
    pdftitle={Symmetries, insulators, and superconductivity in continuum-model of twisted WSe2},    
    pdfauthor={Christos, Bonetti, Scheurer},     
    pdfsubject={},   
    pdfcreator={},   
    pdfproducer={}, 
    pdfkeywords={} {} {}, 
    pdfnewwindow=true,      
    colorlinks=true,       
    linkcolor=blue, 
    citecolor=blue,        
    filecolor=magenta,      
    urlcolor=blue           
}

\newcommand{\equref}[1]{Eq.~(\ref{#1})}

\newcommand{\figref}[1]{Fig.~\ref{#1}}
\newcommand{\refcite}[1]{Ref.~\onlinecite{#1}}

\newcommand{\tableref}[1]{Table~\ref{#1}}

\newcommand{\pdagger}{{\phantom{\dagger}}}

\renewcommand{\approx}{\simeq}

\renewcommand{\vec}[1]{\boldsymbol{#1}}

\definecolor{bleudefrance}{rgb}{0.19, 0.55, 0.91}

\definecolor{wrongultramarine}{rgb}{1,0.5,0}

\begin{document}

\title{Approximate symmetries, insulators, and superconductivity \\ in continuum-model description of twisted WSe$_2$}

\author{Maine Christos}
\affiliation{Department of Physics, Harvard University, Cambridge MA 02138, USA}

\author{Pietro M.~Bonetti}
\affiliation{Department of Physics, Harvard University, Cambridge MA 02138, USA}

\author{Mathias S.~Scheurer}
\affiliation{Institute for Theoretical Physics III, University of Stuttgart, 70550 Stuttgart, Germany}

\begin{abstract}
Motivated by the recent discovery of superconductivity in twisted bilayer WSe$_2$, we analyze the correlated physics in this system in the framework of a continuum model for the moiré superlattice. 
Using the symmetries in a fine-tuned limit of the system, we identify the strong-coupling ground states and their fate when the perturbations caused by finite bandwidth, displacement field, and the phase of the intralayer potential are taken into account. We classify the superconducting instabilities and, employing a spin-fermion-like model, study the superconducting instabilities in proximity to these insulating particle-hole orders. This reveals that only a neighboring intervalley coherent phase (with zero or finite wave vector) is naturally consistent with the observed superconducting state, which we show to be crucially affected by the non-trivial band topology. Depending on details, the superconductor will be nodal or a chiral gapped state while further including electron-phonon coupling leads to a fully gapped, time-reversal symmetric pairing state. 
\end{abstract}

\maketitle

The observation of superconductivity in twisted bilayer graphene \cite{SCTBG} has sparked enormous interest, ultimately making the field of twisted van der Waals moiré superlattices one of the most active areas of current condensed matter research \cite{andreiGrapheneBilayersTwist2020, balentsSuperconductivityStrongCorrelations2020}. While  tight-binding model descriptions on the moiré scale have been discussed and developed early on \cite{PhysRevX.8.031088,PhysRevX.8.031087,PhysRevB.99.195455}, the continuum model \cite{dos2007graphene,MeleModel,bistritzer2011moire,dos2012continuum}, supplemented by interactions, has remained the central model studied by theorists; important insights and guidance have been provided by unrestricted mean-field and analytical strong-coupling studies \cite{PhysRevX.10.031034,PhysRevLett.124.097601,PhysRevB.102.035136,PhysRevB.104.115167,PhysRevResearch.3.013033,PhysRevB.103.205414,PhysRevX.11.041063,PhysRevX.12.021018,PhysRevLett.128.156401} of continuum models. Recent experiments \cite{TunnelingYazdani,TunnelingPerge,JIAsTrilayerScreening,DiodeEffect,chen2023strong,TIVC,2024arXiv240409909Z} on twisted graphene systems yield crucial constraints for superconductivity and point towards an interesting pairing state and mechanism, the nature of which is still under debate \cite{PhysRevB.106.104506,Scammell_2022,PhysRevLett.130.216002,PhysRevB.107.L020502,PhysRevLett.131.016003,BandoffdiagonalPairing,2023arXiv230315551L,VestigialPairing}.

On top of previous reports of insulating behavior and symmetry breaking in twisted bilayer WSe$_2$ \cite{WSe2Ins2,WSe2Ins1,2024arXiv240603315K} as well as indications of superconductivity \cite{WSe2Ins2}, very recently, a genuine superconducting phase was observed \cite{Experiment1,Experiment2}. Most remarkably, the superconducting state appears close to the point where the insulator is suppressed, hinting at an electronic pairing mechanism. While there have been multiple insightful theoretical studies \cite{2021arXiv211010172S,FuMagicAngle,PhysRevResearch.2.033087,PhysRevResearch.4.043048,PhysRevB.104.075150,CiaranSLs,PhysRevResearch.5.L012034,PhysRevB.100.060506,PhysRevB.106.235135,PhysRevB.108.L201110,PhysRevB.108.155111,PhysRevB.108.064506,PhysRevLett.130.126001,2024arXiv240305903A} of the moiré Hubbard model of twisted bilayer WSe$_2$ \cite{PhysRevLett.122.086402}, of extensions thereof, as well as of its strong-coupling spin models, and of patch models \cite{PhysRevLett.130.126001,PhysRevB.104.195134}, we here complement these works by studying the correlated physics directly in a continuum description of the moiré superlattice. We perform a strong-coupling analysis for the insulators, classify the possible superconducting phases and study the resulting pairing states stabilized by fluctuations of the neighboring insulators. Our primary focus is on the symmetries and energetics of these phases and not on the possibility of a direct transition between the two, which could be explained by a spin liquid \cite{2024arXiv240603525K}. We hope that our work can help elucidate the interplay of insulating and superconducting physics in the continuum model of twisted bilayer WSe$_2$, which combined with future experiments, could also help resolve related open questions in graphene-based moiré systems.

\textit{Model and symmetries.---}For the non-interacting physics, we start from the continuum model \cite{PhysRevLett.122.086402,PhysRevResearch.2.033087,FuMagicAngle} $H_{\text{c}} = \int_{\vec{r}} \sum_{\eta=\pm} \sum_{\ell,\ell'=\pm} c^\dagger_{\eta,\ell}(\vec{r}) h^\eta_{\ell,\ell'}(\vec{\nabla},\vec{r}) c^\pdagger_{\eta,\ell'}(\vec{r})$ where
\begin{align}
    h^+ &= - \frac{(- i \vec{\nabla}- \ell_z \vec{\kappa})^2}{2m} + \frac{D}{2}\ell_z + 2 V \hspace{-0.4em} \sum_{j=1,3,5} \cos(\vec{g}_j\vec{r} + \ell_z \psi) \nonumber \\ &\quad  + \left[ \ell_+ w\left(1 + e^{-i \vec{g}_2\vec{r}} + e^{-i \vec{g}_3\vec{r}}\right)/2 + \text{H.c.} \right] \label{Model}\end{align}
describes the states in valley K (spin-$\uparrow$). 
In \equref{Model}, $D$ is the displacement field, $\ell_j$ are Pauli matrices in layer space, $\ell_+ = \ell_x+i\ell_y$, $\vec{g}_j = R(\pi(j-1)/3) \vec{g}_1$, with $R(\varphi)$ rotating 2D vectors by $\varphi$, $\vec{g}_{1,2}$ are the basis vectors of the reciprocal lattice (RL) on the moiré scale, and $\vec{\kappa}$ is the K-point in the single layer Brillouin zone. The Hamiltonian in the other valley (spin-$\downarrow$) is just related by time-reversal symmetry, $h^-_{\ell,\ell'} = (h^+_{\ell,\ell'})^*$. 

\begin{figure}[tb]
   \centering
    \includegraphics[width=\linewidth]{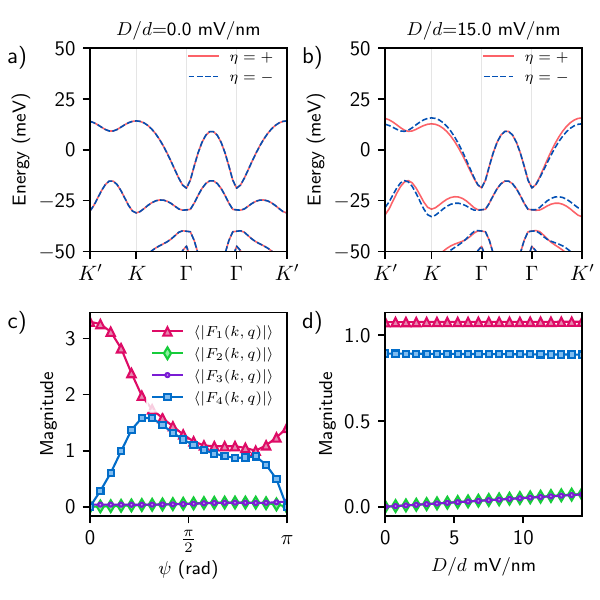}
    \caption{Band structure of the continuum model in \equref{Model} at (a) $D/d=0$ mV/nm and (b) $D/d=15$ mV/nm. We also plot the four components of the form factors in \equref{FormOfFormFactors} as a function of (c) $\psi$ for $D/d=14.25$ mV/nm and (d) as a function of $D$ for $\psi = 128^\circ$. Here $d \approx 0.26 \textrm{\,e}\cdot \textrm{nm}$ is the dipole-moment relating $D$ to the electric field \cite{Experiment1}. All other model parameters are the same as those in \cite{FuMagicAngle} unless otherwise specified.}
    \label{fig:bandstructure}
\end{figure}

Apart from three-fold rotational symmetry, $C_{3z}$, reflection, $m_{x}$, along the $xz$-plane, 
U(1)-valley symmetry, U(1)$_v:\, c(\vec{r}) \rightarrow e^{i\eta_z \varphi} c(\vec{r})$, and time-reversal symmetry $\Theta = i\eta_y \mathcal{K}$ \footnote{Since SU(2) spin rotation invariance is broken, one could also choose $\Theta = \eta_x \mathcal{K}$, cf.~SI of \refcite{BandoffdiagonalPairing}; most notably, this would effectively swap $A_1$ and $A_2$ in \tableref{PairingStates}.}, where $\eta_j$ are Pauli matrices in valley/spin space and $\mathcal{K}$ denotes complex conjugation, the continuum model (\ref{Model}) exhibits the following symmetries: if there is no displacement field, $D=0$, the Hamiltonian is invariant under the (unitary) ``intravalley inversion symmetry'' $\mathcal{I}:\,c(\vec{r}) \rightarrow \ell_x c(-\vec{r})$.
As it will help us understand the correlated physics later on, we will also consider the limit $\psi \in \pi \mathbbm{Z}$ yielding  the additional ``anti-unitary $C_{2z}$ symmetry'' $\mathcal{C}_2$ with action $\,c(\vec{r}) \rightarrow c(-\vec{r})$ (and additional complex conjugation). 

Since we are interested in the low-energy physics, we will only include the single band per valley that intersects the chemical potential and, thus, leads to Fermi surfaces. Denoting the associated band energies by $E_{\vec{k},\eta} =  \epsilon_{\vec{k}} + \eta \, \delta \epsilon_{\vec{k}}$, $\Theta$ implies $\epsilon_{\vec{k}} = \epsilon_{-\vec{k}}$ and $\delta \epsilon_{\vec{k}} = -\delta \epsilon_{-\vec{k}}$. Meanwhile $\mathcal{C}_2$ has no consequences for the band energies (only for the wave functions to be discussed below) and $\mathcal{I}$, if present, enforces $\delta \epsilon_{\vec{k}} = 0$, i.e., the band energies will be even in $\vec{k}$ and identical in the two valleys. These features are clearly visible in our band calculations shown in \figref{fig:bandstructure}(a,b). Denoting the associated electronic annihilation operators in those bands by $d_{\vec{k},\eta}$, we summarize all symmetries in \tableref{SymmetriesAndReps}.

To include interaction effects, we add the Coulomb interaction, also projected to the two bands at the Fermi level, yielding the total Hamiltonian $H = H_0 + H_1$, where the non-interacting and interacting parts are given by $H_0 = \sum_{\vec{k},\eta} E_{\vec{k},\eta} d^\dagger_{\vec{k},\eta} d^\pdagger_{\vec{k},\eta}$ and $H_1 = \frac{1}{2N} \sum_{\vec{q}} V(\vec{q}) \rho_{\vec{q}} \rho_{-\vec{q}}$, respectively, with density operator $\rho_{\vec{q}} = \sum_{\vec{k},\eta} F_{\eta}(\vec{k},\vec{q}) d^\dagger_{\vec{k}+\vec{q},\eta} d^\pdagger_{\vec{k},\eta}$. 
At this point, the precise form of $V(\vec{q})$ is not important; we only assume that $V(\vec{q}) = V(-\vec{q}) \geq 0$ (repulsive). We further already used U(1)$_v$ to constrain the form factors, $F_{\eta}(\vec{k},\vec{q})$, to be diagonal in the valley index. Let us expand these form factors using real coefficients $F_j$,\begin{equation}
    F_{\eta}(\vec{k},\vec{q}) = F_{1}(\vec{k},\vec{q}) + i F_{2}(\vec{k},\vec{q}) + \eta F_{3}(\vec{k},\vec{q}) + i \eta F_{4}(\vec{k},\vec{q}), \label{FormOfFormFactors}
\end{equation}
where $\Theta$ implies further 
\begin{equation}
    F_{1,4}(\vec{k},\vec{q}) = F_{1,4}(-\vec{k},-\vec{q}), \, F_{2,3}(\vec{k},\vec{q}) = - F_{2,3}(-\vec{k},-\vec{q}). \label{TRSFFs}
\end{equation}
At $D=0$, the additional $\mathcal{I}$ symmetry implies that $F_2 = F_3 = 0$. If we further set $\psi \in \pi \mathbb{Z}$, also $F_4$ has to vanish as a result of $\mathcal{C}_2$ and we are simply left with $\rho_{\vec{q}} = \sum_{\vec{k},\eta} F_1(\vec{k},\vec{q}) d^\dagger_{\vec{k}+\vec{q},\eta} d^\pdagger_{\vec{k},\eta}$. We have verified these conclusions by diagonalization of the continuum model, as illustrated in \figref{fig:bandstructure}(c,d).
In combination with $\mathcal{I}$ implying $\delta \epsilon_{\vec{k}} =0$, the Hamiltonian is invariant under an arbitrary, momentum independent SU(2) rotation in valley space, $d_{\vec{k}} \rightarrow U d_{\vec{k}}$, $U = e^{i \vec{\varphi} \cdot \eta_j}$, for $D=0$ and $\psi \in \pi \mathbb{Z}$.

\begin{table}[tb]
\begin{center}
\caption{Symmetries $S$ (including redundant combinations for clarity, see text for more details), their representation on the operators in the band space, when they are present, and the most relevant consequences for our analysis.}
\label{SymmetriesAndReps}\begin{ruledtabular}\begin{tabular}{ccccc} 
$S$ & unitary & $S d_{\vec{k}} S^{-1}$ & Condition & Consequences \\ \hline
U(1)$_v$ & \cmark & $e^{i\varphi \eta_z}d_{\vec{k}}$ & --- & $F_{\eta,\eta'} \propto \delta_{\eta,\eta'}$   \\
$\Theta$ & \xmark & $i\eta_y d_{-\vec{k}}$ & --- & $E_{\vec{k},\eta}=E_{-\vec{k},-\eta}$, \equref{TRSFFs} \\
$C_{3z}$ & \cmark & $d_{C_{3z}\vec{k}}$ & --- & $E_{\vec{k},\eta}=E_{C_{3z}\vec{k},\eta}$ \\
$m_{x}$ & \cmark & $i\eta_y d_{m_{x} \vec{k}}$ & --- & $E_{\vec{k},\eta}=E_{m_{x} \vec{k},-\eta}$ \\
$\Theta m_{x}$ & \xmark & $-d_{-m_{x} \vec{k}}$ & --- & $E_{\vec{k},\eta}=E_{-m_{x} \vec{k},\eta}$ \\
$\mathcal{I}$ & \cmark & $d_{-\vec{k}}$ & $D=0$ & $E_{\vec{k},\eta}=E_{-\vec{k},\eta}$ \\
$\Theta\mathcal{I}$ & \xmark & $ i\eta_yd_{\vec{k}}$ & $D=0$ & $F_2 = F_3 =0$ \\
$\mathcal{C}_2$ & \xmark & $d_{\vec{k}}$ & $\psi \in \pi \mathbbm{Z}$ & $F_2 = F_4 = 0$
 \end{tabular}
\end{ruledtabular}
\end{center}
\end{table}

\textit{Correlated insulators.---}To study interacting ground states at half filling, we start from this simple limit and further set the bandwidth $W$ to zero. Consider the family of product states as candidates for $\nu=1$ 
\begin{equation}
    \ket{\hat{\vec{n}}} : = \prod_{\vec{k}} \left( d^\dagger_{\vec{k}}  U^\dagger_{\hat{\vec{n}}}\right)_+ \ket{0}, \, U^\dagger_{\hat{\vec{n}}} \eta_z U^\pdagger_{\hat{\vec{n}}} = \hat{\vec{n}} \cdot \vec{\eta}, \label{Statesn}
\end{equation}
where $\hat{\vec{n}}$ is a three-component unit vector. 
Due to the simple form $\rho_{\vec{q}} = \sum_{\vec{k},\eta} F_1(\vec{k},\vec{q}) d^\dagger_{\vec{k}+\vec{q},\eta} d^\pdagger_{\vec{k},\eta}$ of the density operator, we find 
\begin{equation}
    \rho_{\vec{q}} \ket{\hat{\vec{n}}} = \sum_{\vec{G} \in \text{RL}} \delta_{\vec{q},\vec{G}} f_{\vec{G}} \ket{\hat{\vec{n}}}, \label{RhoOnState}
\end{equation}
with $f_{\vec{G}} = \sum_{\vec{k}} F_1(\vec{k},\vec{G})$, i.e., the state is annihilated by the shifted density operator $\bar{\rho}_{\vec{q}} := \rho_{\vec{q}} - \sum_{\vec{G}} \delta_{\vec{q},\vec{G}} f_{\vec{G}}$. Upon writing $H_1 = \bar{H}_1 + \Delta H_1$ with $\bar{H}_1 = \frac{1}{2N} \sum_{\vec{q}} V(\vec{q}) \bar{\rho}_{\vec{q}}^\dagger \bar{\rho}_{\vec{q}}$ and $\Delta H_1 = \sum_{\vec{G}} V(\vec{G}) f_{\vec{G}} ( 2\rho_{-\vec{G}} - f_{-\vec{G}})$.
If we further assume the ``flat-metric condition'' \cite{PhysRevB.103.205411}, i.e., that $F_1(\vec{k},\vec{G})$ is independent of $\vec{k}$, which we verified holds approximately for the model (\ref{Model}) \cite{SI}, we get that $\rho_{\vec{G}}$ is proportional to the particle-number operator; in that case, $\Delta H_1$ is just a constant at fixed filling. Since $\bar{H}_1 \ket{\hat{\vec{n}}} = 0$ by virtue of \equref{RhoOnState} and noting that $\bar{H}_1$ is positive semi-definite, we conclude that any $\ket{\hat{\vec{n}}}$ in \equref{Statesn} is an exact ground state of the interacting part of the Hamiltonian. 

When also including the dispersion part of the Hamiltonian $H_0$, these states still remain exact, degenerate eigenstates of the full Hamiltonian for $D=0$ and $\psi \in \pi \mathbb{Z}$. Note, however, that they will cease to be the ground state above a certain value of the bandwidth. While eventually, the system is expected to become a symmetry unbroken metal at large bandwidth, there might also be an intermediate regime where translational symmetry breaking order develops, as seen in  triangular lattice Hubbard models \cite{FuMagicAngle,PhysRevResearch.2.033087,PhysRevResearch.4.043048,PhysRevB.104.075150,CiaranSLs,PhysRevB.108.L201110,PhysRevB.108.155111,PhysRevB.108.064506,PhysRevB.98.075109,PhysRevB.40.2727,PhysRevLett.69.2590,Deutscher1993}. 

\textit{Dispersion as perturbation.---}We next analyze what happens when the SU(2) symmetry is broken due to finite $\delta \epsilon_{\vec{k}}$ in the dispersion, i.e., when $D$ is turned on. It is straightforward to see that, within first order perturbation theory, there is no energetic splitting between the states in \equref{Statesn}, which is a consequence of $\sum_{\vec{k}} \delta \epsilon_{\vec{k}} = 0$ (imposed by $\Theta$). Since $\eta_z$ ($\eta_{x,y}$) commutes (anti-commute) with the associated perturbation $\eta_z \delta \epsilon_{\vec{k}}$, its second order contributions is still zero (finite) \cite{PhysRevX.12.021018}; this can also be directly seen by noting that $\ket{\hat{\vec{e}}_{z}}$ is still an exact eigenstate of the full Hamiltonian while $\ket{\hat{\vec{e}}_{x,y}}$ are not. Since corrections to the ground state energy in second order of perturbation theory are always non-positive, this shows that the intervalley coherent (IVC) states $\ket{\hat{\vec{e}}_{x,y}}$ are energetically favored over the valley  polarized (VP) state $\ket{\hat{\vec{e}}_{z}}$. As such, superexchange processes controlled by $D$ break the SU(2) symmetry in a way that leads to an ``easy plane'' for $\hat{\vec{n}}$.

\textit{Form factors as perturbation.---}Another important perturbation to the SU(2) manifold comes from the form factors. As discussed above, turning on $D$ at $\psi \in \pi \mathbbm{Z}$ allows for finite $F_3$ in \equref{FormOfFormFactors}, while $\psi \notin \pi \mathbbm{Z}$ at $D=0$ admixes $F_4$ to the $F_1$ component. Finally, when both $\psi \notin \pi \mathbbm{Z}$ and $D\neq 0$ also $F_2$ can become finite. 
To analyze which state is favored by these perturbations, we compute the change $\Delta E(\hat{\vec{n}})$ of $\braket{\hat{\vec{n}}|H_1|\hat{\vec{n}}}$ associated with the form factors $F_{j>1}$; using $F_{j}(\vec{k},\vec{G}) = s_j F_{j}(-\vec{k},\vec{G})$ with $s_{1,2}=-s_{3,4}=1$, as follows from $\Theta$ and Hermiticity, we obtain
\begin{equation}
    \Delta E(\hat{\vec{n}}) = \Delta E_0 - \sum_{\vec{k},\vec{q}} \frac{V(\vec{q})}{2N} \left[F_{3}^2(\vec{k},\vec{q}) + F_{4}^2(\vec{k},\vec{q})\right] \hat{n}^2_z,
\end{equation}
where $\Delta E_0$ is an SU(2)-invariant and, thus, $\hat{\vec{n}}$-independent contribution. Therefore, both the $\psi$- and $D$-induced corrections to the form factors lead to an ``easy axis'' for $\hat{\vec{n}}$, favoring the VP state. This also holds beyond this simple variational calculation (if the flat-metric condition holds) by noting that \equref{RhoOnState} still applies for $\ket{\pm \hat{\vec{e}}_z}$ (albeit with a modified $f_{\vec{G}} = \sum_{\vec{k}} [F_1(\vec{k},\vec{G}) + i F_2(\vec{k},\vec{G})]$), but not for $\ket{\pm \hat{\vec{e}}_{x,y}}$, establishing VP as the ground state when $\psi \notin \pi \mathbbm{Z}$ and/or $D\neq 0$ in the flat-band limit.

Taken together, depending on whether the $D$-induced SU(2)-symmetry breaking in the band structure or the additional form factors induced by $D$ and $\psi$ are larger, we get an IVC or VP ground state. At $D=0$, the VP ground state is favored for generic values of $\psi$. As indicated by experiment, additional corrections to this description, e.g., coming from thermal fluctuations in the vicinity of the SU(2) symmetric point, the finite bandwidth and additional corrections to the interaction, the system ceases to be an interaction-induced insulator and, instead, develops superconductivity for small $D$.

\begin{table}[tb]
\begin{center}
\caption{Possible pairing states classified according to the IRs of the point group $C_{3v}$ of the continuum model. We list the minimal number of nodes on a Fermi surface encircling the $\Gamma$ point with the three values referring to first $D\neq 0$ (and sufficiently large, see text), second $D=0$ and $\mathcal{I}$ even, and, third, $D=0$ and $\mathcal{I}$ odd, respectively. As a result of $\Theta$ in the normal state, all basis functions are real. Furthermore, $\chi_{\vec{k}}$ and $(X_{\vec{k}},Y_{\vec{k}})$ transform as a scalar and vector under $C_{3z}$, respectively, and $X_{\vec{k}} = X_{-m_{x} \vec{k}}$, $Y_{\vec{k}} = -Y_{-m_{x} \vec{k}}$. In the last column, we list the additional two possible pairing states associated with $\mathcal{I}$ at $D=0$.}
\label{PairingStates}\begin{ruledtabular}\begin{tabular}{ccccc} 
IR of $C_{3v}$ & form of $\Delta_{\vec{k}}$ & nodes & for $D=0$  \\ \hline
$A_1$ & $\chi_{\vec{k}} = \chi_{-m_{x} \vec{k}}$ & $0/0/6$ & $\chi_{\vec{k}} = \pm \chi_{-\vec{k}} = \pm \chi_{m_{x} \vec{k}}$ \\
$A_2$ & $\chi_{\vec{k}} = -\chi_{-m_{x} \vec{k}}$ & $6/12/6$ & $\chi_{\vec{k}} = \pm \chi_{-\vec{k}} = \mp \chi_{m_{x} \vec{k}}$ \\
\begin{tabular}[c]{@{}c@{}c@{}}$E(1,0)$ \\$E(0,1)$ \\ $E(1,i)$ \end{tabular} & \begin{tabular}[c]{@{}c@{}c@{}}$X_{\vec{k}}$ \\ $Y_{\vec{k}}$ \\ $X_{\vec{k}}+i Y_{\vec{k}}$ \end{tabular} & \begin{tabular}[c]{@{}c@{}c@{}}$2/4/2$ \\$2/4/2$ \\ $0/0/0$ \end{tabular} & \begin{tabular}[c]{@{}c@{}} $X_{\vec{k}} = \pm X_{-\vec{k}} = \pm X_{m_{x} \vec{k}}$ \\$X_{\vec{k}} = \pm Y_{-\vec{k}} = \mp Y_{m_{x} \vec{k}}$  \end{tabular}
 \end{tabular}
\end{ruledtabular}
\end{center}
\end{table}

\textit{Classification of pairing.---}Before we study the energetics of superconductivity, we use the symmetries in \tableref{SymmetriesAndReps} to classify the different forms of pairing instabilities. As a result of U(1)$_v$, pairing is either entirely intra- or intervalley and we focus on the latter here as there are no indications \cite{Experiment1,Experiment2} that the normal state above $T_c$ has broken time-reversal symmetry. 
On the mean-field level, the superconducting order parameter couples as $H_{\text{MF}} = \sum_{\vec{k},\eta} d^\dagger_{\vec{k},\eta} \Delta_{\vec{k},\eta} d^\dagger_{-\vec{k},-\eta}$ to the low-energy electrons. Fermi-Dirac statistics implies $\Delta_{\vec{k},\eta} = -\Delta_{-\vec{k},-\eta}$ such that we will focus on $\Delta_{\vec{k}} := \Delta_{\vec{k},+}$ in the following without loss of generality.

Using the irreducible representations (IRs) of the point group $C_{3v}$, we find five different pairing instabilities, as summarized in \tableref{PairingStates}---two associated with the one-dimensional IRs, $A_1$, $A_2$, and three for the two-dimensional IR $E$. We emphasize that, by virtue of the intervalley nature of $m_{x}$, its action on the superconducting order parameter reads as $\Delta_{\vec{k}} \rightarrow \Delta_{-m_{x} \vec{k}}$.  
Importantly, at $D=0$, the additional intravalley inversion symmetry $\mathcal{I}$ leads to a splitting of each of the above-mentioned IRs $g \in \{A_1,A_2,E\}$ into two, which we denote by $g^\pm$ and are associated with the two possible signs $\pm$ in the constraint $\Delta_{\vec{k}} = \pm \Delta_{-\vec{k}}$. This is indicated in the last column of \tableref{PairingStates}. Depending on the sign, $\mathcal{I}$ does or does not give rise to additional symmetry-protected nodes. By continuity, these additional nodes are stable against finite $D$.

\begin{figure}[tb]
   \centering
    \includegraphics[width=\linewidth]{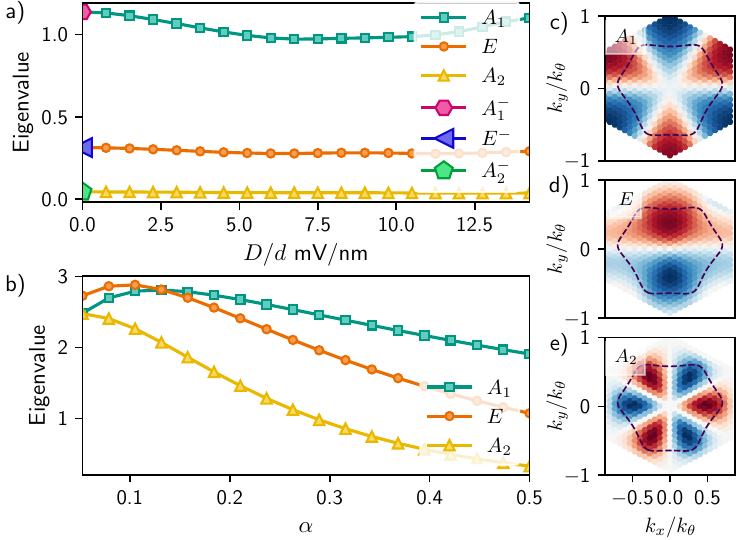}
    \caption{Largest eigenvalues of the LGE as a function of (a) displacement field $D$ and (b) the mass parameter $\alpha$ of the fluctuating bosons corresponding to the IRs $A_1$ ($A_1^-$), $A_2$ ($A_2^-$), and $E$ ($E^-$) for $D > 0$ ($D=0$). The order parameter $\Delta_{\vec{k}}$ of the leading, subleading, and third leading states in (a) at $D/d=14.25$ mV/nm are shown in (c), (d), and (e) respectively. }
    \label{fig:SCConnectedToStrong}
\end{figure}

\textit{Energetics of superconductivity.---}As in both experiments \cite{Experiment1,Experiment2} the superconducting state appears in close proximity to the insulating phase, we first assume that fluctuations of the latter provide the pairing glue. Therefore, we study, in analogy to the celebrated spin-fermion model \cite{spinfermion_model} for the high-temperature superconductors, an effective model, where the low-energy fermions $d_{\vec{k}}$ are coupled to collective bosons $\phi_j(\vec{q}) = \phi_j^\dagger(-\vec{q})$ via
\begin{equation}
    H_{\text{fb}} = \sum_{\vec{k},\vec{q}} \sum_{j=x,y,z} d^\dagger_{\vec{k}+\vec{q}} \lambda_j(\vec{k},\vec{q}) d^\pdagger_{\vec{k}} \phi_j(\vec{q}). \label{SFInteraction}
\end{equation}
Here $j=x,y$ describe IVC and $j=z$ valley fluctuations constraining their associated form factors $\lambda_{x,y}$ and $\lambda_z$ to be valley off-diagonal and diagonal, respectively, while time-reversal implies $\eta_y \lambda_j^*(\vec{k},\vec{q}) \eta_y = - \lambda_j(-\vec{k},-\vec{q})$.

Inspired by the strong-coupling analysis above, we start by assuming that the (static) susceptibility $\chi(\vec{q})$ of the collective bosons is peaked at $\vec{q}=0$. The boson $\phi_z$ will mediate a repulsive intra-valley Cooper channel interaction and is, thus, not expected to stabilize superconductivity. To demonstrate that, we solve the linearized gap equations (LGE) numerically, using $\chi(\vec{q}) = C\frac{\alpha}{\alpha^2+\vec{q}^2/k_\theta^2}$ and taking $\lambda_j$ in \equref{SFInteraction} to be the projections of $\eta_j$ 
onto the Bloch states of the band at the Fermi level \cite{SI}; here $\alpha$ parametrizes the bosonic mass which becomes smaller when approaching the neighboring insulating phase and we take $\alpha=1$ and $C/A_{\text{moire}}=1.7$ meV unless otherwise specified. Indeed, we find vanishingly small eigenvalues for the case of fluctuations of $\phi_z$. This is to be contrasted with fluctuations of $\phi_{x,y}$, which mediate an attractive intervalley interaction, in the sense that they favor $\Delta_{\vec{k},\eta} \approx \Delta_{\vec{k},-\eta}$. However, due to the Fermi-Dirac constraint $\Delta_{\vec{k},\eta} = -\Delta_{-\vec{k},-\eta}$ and the significantly larger band separation compared to twisted bi- and trilayer graphene, where interband pairing is stabilized by these types of interactions \cite{BandoffdiagonalPairing}, we here have $\Delta_{\vec{k}} \approx -\Delta_{-\vec{k}}$ and, thus, sign changes of the order parameter in the Brillouin zone. While this indicates that the order parameter transforms under an $\mathcal{I}$-odd representation for $D\rightarrow 0$, all of $A_1^-$, $A_2^-$, and $E^-$ are in principle possible and the outcome depends on energetics. In line with these considerations, we find in our numerics in \figref{fig:SCConnectedToStrong}(a,b) that the states with largest pairing strength (eigenvalue) not only emanate from $\mathcal{I}$-odd IRs [see (a)] but also stay approximately odd in $\vec{k}$ in the experimentally relevant range of $D$, see \figref{fig:SCConnectedToStrong}(c-e). 
We further observe that the $A_1^-$ and $E^-$ states are competing and which of the two is favored depends on the proximity to the insulator (the value of $\alpha$). We note that the dominance of $A_1$ pairing for larger $\alpha$ is crucially related to the non-zero Chern number of the bands, which requires $\lambda_j(\vec{k},\vec{q})$ to vanish at $\vec{k}=\vec{q}=0$ \cite{SI}.
As a result of the additional symmetry-imposed lines of zeros of $\Delta_{\vec{k}}$ at the zone boundary, the $A_2^-$ state in \figref{fig:SCConnectedToStrong}(e) has very small eigenvalues.

\begin{figure}[tb]
   \centering
    \includegraphics[width=\linewidth]{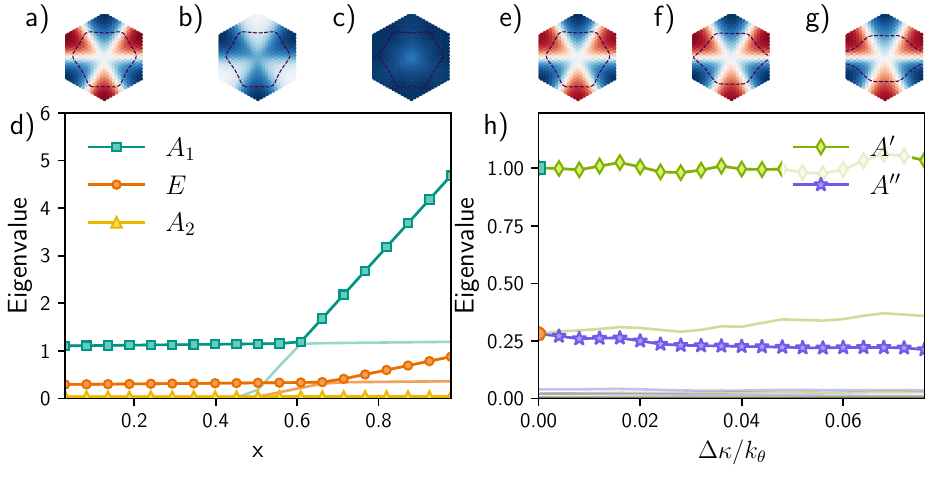}
    \caption{Tuning the relative strength of the pairing interactions from purely IVC fluctuations ($x=0$) to only electron-phonon coupling ($x=1$), we show $\Delta_{\vec{k}}$ of the leading state for (a) $x=0$, (b) $x=0.61$, (c) $x=1$, and the evolution of the eigenvalues of the LGE in at $D/d=14.25$ mV/nm (d); here lines with the same color correspond to the same IR of $C_{3v}$. (e-h)~show the same quantities but as a function of $\Delta\kappa$, capturing the $C_{3z}$ breaking due to strain or nematic order, and using the IRs of $C_{s}$ at $D/d=5$ mV/nm. 
    The plot markers at $\Delta \kappa = 0$ refer to the legend in (d). 
    }
    \label{fig:SCExtended}
\end{figure}

As noted above, the significant value of the dispersion can lead to ordering at finite wave vector; this is why we also considered $\chi(\vec{q})$ that are peaked at three, $C_{3z}$ related, minima $(C_{3z})^{j-1} \vec{q}_f$ with $\vec{q}_f$ being varied from $\vec{q}_f=0$ to the K point. We find \cite{SI} that the above conclusions about the order of eigenvalues is not affected until $\vec{q}_f$ gets close to the K point, where $A_2$ can become the leading state; its eigenvalue is, however, comparatively small.

Motivated by indications that the electron-phonon coupling in monolayer transition metal dichalcogenides can become sizable \cite{PhysRevX.9.031019}, we here study the additional impact of phonons on pairing, focusing on $A_1'$ optical phonons of the constituent monolayers \cite{PhysRevB.84.155413}, seen in Raman studies \cite{RamanMonolayer,PhysRevX.9.031019}.
These can mediate an attractive intravalley interaction \cite{SI} and are, thus, expected to favor the conventional $A_1^+$ state. This can also be seen in our numerics in \figref{fig:SCExtended}(a-d), where we find a transition (crossover at $D\neq 0$) from (primarily) $A_1^-$ to $A_1^+$, when phonon-mediated interactions become comparable to those induced by IVC fluctuations. We here introduced the dimensionless parameter $x\in[0,1]$ parameterizing the ratio $x/(1-x)$ of electron-phonon- to IVC-fluctuation-induced interactions \cite{SI}.

We finally discuss the impact of rotational symmetry breaking, as it can result from strain \cite{PhysRevB.100.035448,StrainWSe2} or possibly electronic nematic order. We capture this symmetry reduction by the shift of magnitude $\Delta \kappa$ of $\vec{\kappa}$ in \equref{Model}; choosing for concreteness a high symmetry direction of this shift that retains the reflection symmetry $m_x$, the point group (at finite $D$) becomes $C_s$.
We can see in \figref{fig:SCExtended}(e-g) that the states associated with IR $E$ split into two---one even (IR $A'$ of $C_s$) and one odd ($A''$) under $m_x$. While the leading state continues to be the previous $A_1$ state (now $A'$), we observe that strain can, for very large values, reduce the number of nodal points [see (g)].

\textit{Conclusion.---}In summary, we have studied the possible insulating phases as a result of Coulomb interaction in a continuum-model description of twisted WSe$_2$, showing that IVC or VP is favored depending on whether the $D$-induced SU(2)-breaking perturbations in the bandwidth or in the form factors dominate. Using the valley-fermion model description of the pairing glue in \equref{SFInteraction}, we find that only IVC fluctuations are consistent with superconductivity, necessarily leading to a superconducting order parameter with nodes or broken time-reversal symmetry. As samples with superconductivity show insulating (superconducting) behavior at larger $D$ (smaller $D$), this IVC-based pairing glue is also consistent with the strong-coupling analysis. 
The superconductor becomes fully gapped when additional interactions from phonons are significant. 
There are many natural directions for future works, such as including more bands and generalizing this continuum-model study to heterobilayers \cite{TMDfrg,PhysRevB.102.235423,PhysRevLett.131.056001} or more exotic novel forms of moiré superlattices \cite{2024arXiv240308736W,2024arXiv240412420P}.

\textit{Note added.---}Just before posting our work, \refcite{2024arXiv240619348Z} appeared online, which also discusses pairing in twisted twisted WSe$_2$ in the continuum model.

\begin{acknowledgments}
M.S.S.~acknowledge funding by the European Union (ERC-2021-STG, Project 101040651---SuperCorr). Views and opinions expressed are however those of the authors only and do not necessarily reflect those of the European Union or the European Research Council Executive Agency. Neither the European Union nor the granting authority can be held responsible for them. M.C.~and P.M.B.~acknowledge funding from U.S.~National Science Foundation grant No.~DMR-2245246. P.M.B. acknowledges support by the German National Academy of Sciences Leopoldina through Grant No.~LPDS 2023-06. M.C.~thanks Ming Xie for helpful conversations on pairing mechanisms.
\end{acknowledgments}

\bibliography{Notes_1.bib}

\newpage

\onecolumngrid

\begin{appendix}

\section{Numerical solution of gap equation}\label{NumericsForGapEq}
The most general form of interactions in the space of the active bands we study for pairing is:
\begin{equation}
    H_{\text{att.}}=-\frac{1}{A}\sum_{\vec{k},\vec{k}'}\sum_{\vec{q}} \chi_{\vec{q}}d^\dagger_{\vec{k},\eta}\lambda_j^{\eta\eta'}(\vec{k},\vec{q})d_{\vec{k}+\vec{q},\eta'}d^\dagger_{\vec{k}',\eta''}\lambda_j^{\eta''\eta'''}(\vec{k'},-\vec{q})d_{\vec{k}'-\vec{q},\eta'''} \label{AttractiveInteraction}
\end{equation}
where $A$ is the sample area, $\vec{k}$ and $\vec{k}'$ are summed over the first Brillouin zone and $\vec{q}$ is summed over all momenta. For intervalley fluctuations, i.e., fluctuations of $\phi_{x,y}$, we only keep $j=x,y$ in \equref{AttractiveInteraction}; after using time-reversal symmetry to fix the gauge of the wavefunctions, we find the following LGE:
\begin{equation}\label{GapIVC}
    \left(\Delta^\dagger_{\vec{k}}\right)^{+-}=\sum_{\vec{q}}\left(\chi_{\vec{q}}+\chi_{-\vec{q}}\right)\frac{1-2n_F(E^+_{\vec{k}+\vec{q}})}{2E^+_{\vec{k}+\vec{q}}}\left(\Delta^\dagger_{\vec{k}+\vec{q}}\right)^{-+}|\lambda_{x,y}^{+-}(\vec{k}+\vec{q},-\vec{q})|^2
\end{equation}
 Here $E^+_{\vec{k}+\vec{q}}$ is the dispersion in the $+$ spin-valley flavor with the chemical potential fixed for every value of displacement field $D$ such that hole-doping $\nu=1$. $\Delta^\dagger_{\vec{k}}$ is defined as:
 \begin{equation}
    \left(\Delta_{\vec{k}}\right)^{\eta-\eta}=\sum_{\vec{q}}(\chi_{\vec{q}}+\chi_{-\vec{q}})\lambda_{j}^{\eta\eta'}(\vec{k},\vec{q})\langle d_{-\vec{k}-\vec{q},-\eta'}d_{\vec{k}+\vec{q},\eta'}\rangle \lambda_{j}^{-\eta-\eta'}(-\vec{k},-\vec{q})
 \end{equation}
 In the case where there is more than one type of fluctuation contributing to pairing, the $\lambda_j$ may be summed over $j=x,y,z,0$ in the above. We also multiply the IVC interactions by an additional factor of $\frac{1}{2}$ relative to when we consider fluctuations of an order with only a single component, since at the level of the gap equation, $\lambda_{x}$ and $\lambda_y$ provide an identical contribution. After using Fermi-Dirac statistics, \equref{GapIVC} can be rewritten as:
\begin{equation}
    \left(\Delta^\dagger_{-\vec{k}}\right)^{-+}=-\sum_{\vec{q}}\left(\chi_{\vec{q}}+\chi_{-\vec{q}}\right)\frac{1-2n_F(E^+_{\vec{k}+\vec{q}})}{2E^+_{\vec{k}+\vec{q}}}\left(\Delta^\dagger_{\vec{k}+\vec{q}}\right)^{-+}|\lambda_{x,y}^{+-}(\vec{k}+\vec{q},-\vec{q})|^2 \label{IVCGapEq}
\end{equation}
For VP fluctuations ($\phi_z$) and, thus, $j=z$ in \equref{AttractiveInteraction}, the LGE we solve is:
\begin{equation}
    \left(\Delta^\dagger_{\vec{k}}\right)^{-+}=-\sum_{\vec{q}}\left(\chi_{\vec{q}}+\chi_{-\vec{q}}\right)\frac{1-2n_F(E^+_{\vec{k}+\vec{q}})}{2E^+_{\vec{k}+\vec{q}}}\left(\Delta^\dagger_{\vec{k}+\vec{q}}\right)^{-+}|\lambda_{z}^{++}(\vec{k}+\vec{q},-\vec{q})|^2
\end{equation}
The linearized gap equation for $\phi_0$, or equivalently the electron-phonon interactions we study, is the same as for VP fluctuations but with the opposite sign:
\begin{equation}
    \left(\Delta^\dagger_{\vec{k}}\right)^{-+}=\sum_{\vec{q}}\left(\chi_{\vec{q}}+\chi_{-\vec{q}}\right)\frac{1-2n_F(E^+_{\vec{k}+\vec{q}})}{2E^+_{\vec{k}+\vec{q}}}\left(\Delta^\dagger_{\vec{k}+\vec{q}}\right)^{-+}|\lambda_0^{++}(\vec{k}+\vec{q},-\vec{q})|^2
\end{equation}
Unless otherwise specified, all solutions are calculated at $T=1$ K. We sum $\vec{q}$ out to 2 shells outside the first Brillouin zone, an approximation justified by the falling off of both the form factors and $\chi_{\vec{q}}$.

For the case where we study the relative strength of different types of fluctuations or phonons ($\phi_0$), for example, IVC fluctuations ($\phi_{x,y}$) and phonons [in \figref{fig:SCExtended}(a-d) of the main text], we solve a gap equation of the following form:

\begin{equation}
\begin{split}
    \left(\Delta^\dagger_{\vec{k}}\right)^{-+}=-(1-x)\sum_{\vec{q}}\left(\chi_{\vec{q}}+\chi_{-\vec{q}}\right)&\frac{1-2n_F(E^+_{-\vec{k}-\vec{q}})}{2E^+_{-\vec{k}-\vec{q}}}\left(\Delta^\dagger_{-\vec{k}-\vec{q}}\right)^{-+}|\lambda_{x,y}^{+-}(-\vec{k}-\vec{q},\vec{q})|^2\\&+x\sum_{\vec{q}}\left(\chi_{\vec{q}}+\chi_{-\vec{q}}\right)\frac{1-2n_F(E^+_{\vec{k}+\vec{q}})}{2E^+_{\vec{k}+\vec{q}}}\left(\Delta^\dagger_{\vec{k}+\vec{q}}\right)^{-+}|\lambda_0^{++}(\vec{k}+\vec{q},-\vec{q})|^2
\end{split}
\end{equation}
where $x$ is a parameter which is used to tune the relative strengths of each interactions.

\section{Additional numerical results}\label{NumericalResultsSI}
\begin{figure}
    \centering
    \includegraphics{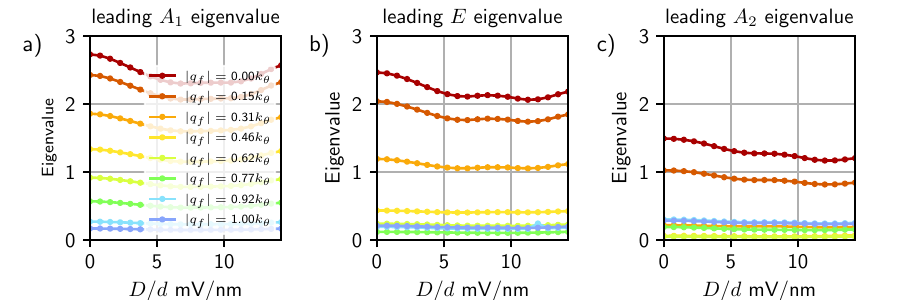}
    \caption{We plot the leading eigenvalues corresponding to the $A_1$ (a), $E$ (b), and $A_2$ (c) representations for varying wave-vectors $\vec{q}_f$ at with IVC fluctuations ($\phi_{x,y}$). We use a  gap parameter $\alpha=0.25$ in $\chi(\vec{q})$.}
    \label{fig:finiteq}
\end{figure}

\begin{figure}
    \centering
    \includegraphics{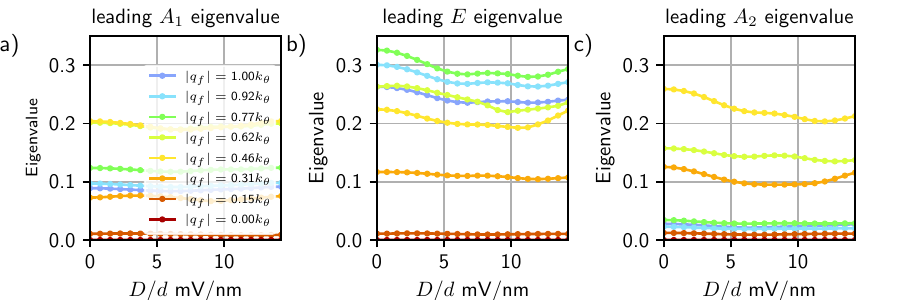}
    \caption{We plot the leading eigenvalues corresponding to the $A_1$ (a), $E$ (b), and $A_2$ (c) representations for varying wave-vectors $\vec{q}_f$ with VP fluctuations $(\phi_z)$. We use a gap parameter $\alpha=0.25$ in $\chi(\vec{q})$. We note the eigenvalues here are significantly smaller than for the case of IVC fluctuations with the same coupling strength and the relative dominance of the $E$ state.}
    \label{fig:finiteqZZ}
\end{figure}

\begin{figure}
    \centering
    \includegraphics{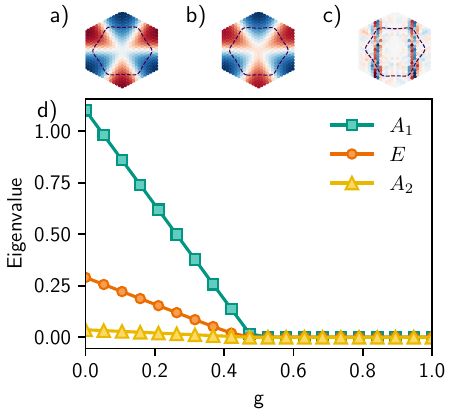}
    \caption{We fix $D/d=14.45$ mV/nm and plot the leading eigenvalue corresponding to relative couplings $g$ between IVC fluctuations and VP fluctuations for $g=0$ (a), and $g=0.42$ (b), and $g=1$ (c), as well as the leading eigenvalues for the $A_1$, $A_2$, and $E$ representations as a function of $g$.}
    \label{fig:ZZ}
\end{figure}

\begin{figure}
    \centering
    \includegraphics{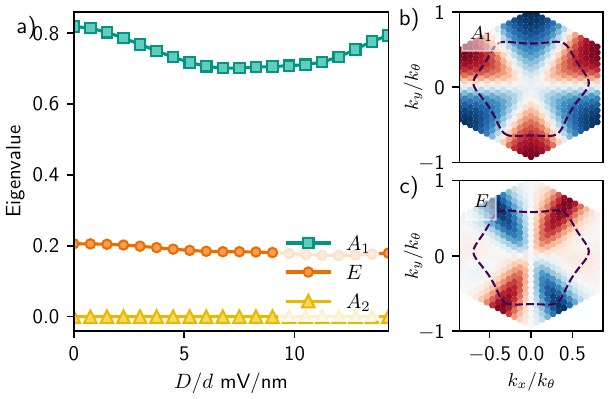}
    \caption{We plot eigenvalue as a function of displacement field (a) as well as the leading (b) and subleading (c) basis functions as a function of displacement field for layer-odd fluctuations of $\phi_{x,y}$.}
    \label{fig:odd}
\end{figure}

\begin{figure}
    \centering
    \includegraphics{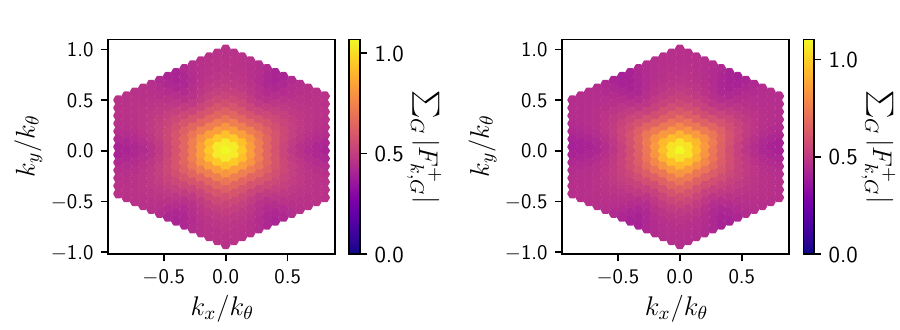}
    \caption{We plot $\sum_{\vec{G}}|F^{+}_{\vec{k},\vec{G}}|$ for the six smallest nonzero $\vec{G}$ reciprocal lattice vectors for $D/d=0$ and $D/d=14.45$ mV/nm.}
    \label{fig:flatmetric}
\end{figure}

\begin{figure}
    \centering
    \includegraphics[width=\linewidth]{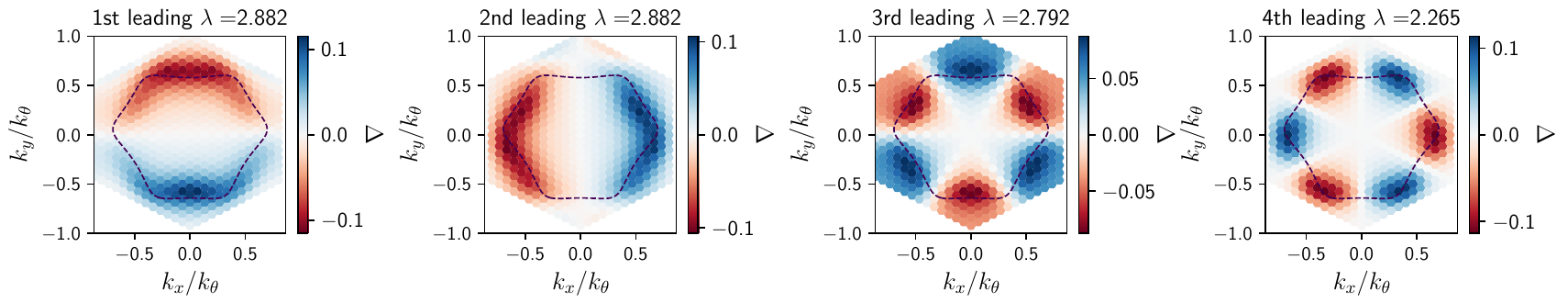}
    \caption{Basis functions at $D/d=14.45$ mV/nm with $\alpha=0.1$ for fluctuations of $\phi_{x,y}$.}
    \label{fig:basis}
\end{figure}

In this supplement, we present solutions of the LGE in the presence of additional perturbations, interactions, or potentials not explicitly shown in the main text. In Fig.~\ref{fig:finiteq}, we show the leading eigenvalues for each IR for a potential $\chi(\vec{q})=\frac{C\alpha}{\alpha^2+(\vec{q}-\vec{q}_f)^2/k_\theta^2}+\frac{C\alpha}{\alpha^2+(\vec{q}-C_{3z}\vec{q}_f)^2/k_\theta^2}+\frac{C\alpha}{\alpha^2+(\vec{q}-C_{3z}^2\vec{q}_f)^2/k_\theta^2}$ with gap parameter $\alpha=0.25$ for different values of $\vec{q}_f$ and IVC fluctuations. We choose $\vec{q}_f$ to lie along the contour between the $\Gamma$ point and $\vec{K}$ point in the moir\'e Brillouin zone. We also present the same type of plots for VP fluctuations in Fig.~\ref{fig:finiteqZZ}.
Furthermore, we illustrate the impact of competition between $\phi_{x,y}$ and $\phi_z$ fluctuations in Fig.~\ref{fig:ZZ}, where as in the case of phonons, we parameterize their relative strengths with a dimensionless parameter, now $g$ instead of $x$, and normalize the contribution from the two component $\phi_{x,y}$ fluctuations by a factor of 2. One can see that VP fluctuations are detrimental to pairing.
In Fig.~\ref{fig:odd}, we study the LGE for layer-odd IVC ($\phi_{x,y}$) fluctuations and demonstrate that our conclusions do not strongly depend on whether the fluctuations are layer even or odd.
What is more, we check the flat metric condition both with and without a nonzero displacement field in Fig.~\ref{fig:flatmetric}. Apart from a peak close to $\Gamma$, the form factors are approximately independent of $\vec{k}$ for $\vec{q}=\vec{G}\in\text{RL}$.
Finally, we show the leading basis functions, which are more sharply peaked around the Fermi surface, for $D/d=14.45$ mV/nm and $\alpha=0.1$ in Fig.~\ref{fig:basis} to demonstrate how a smaller $\alpha$ affects the solutions we find.

\section{Topological constraints on form factors and impact on pairing}
In this appendix, we discuss how the non-zero Chern number of the active bands necessarily leads to a non-trivial momentum dependence of the form factors. As IVC fluctuations provide the most promising pairing glue in our theory, we illustrate this using the intervalley form factors $\lambda^{+-}_{x,y}(\vec{k},\vec{q})$ entering the associated LGE in \equref{IVCGapEq}. Using time-reversal symmetry to relate the wave functions in the two valleys, the form factors can be written as $\lambda^{+-}_{x}(\vec{k},\vec{q}) = \braket{u^*_+(-\vec{k})|u_+(\vec{k}+\vec{q})} $, and similarly for the $y$ component, where $\ket{u_{\eta}(\vec{k})}$ are the Bloch states in valley $\eta$. Viewing $\lambda^{+-}_{x}(\vec{k},\vec{q})$ as a function $s_{\vec{k}_0}(\vec{q})=\lambda^{+-}_{x}(\vec{k}_0,\vec{q})$ of the momentum transfer $\vec{q}$ at fixed $\vec{k}=\vec{k}_0$, we conclude that $s_{\vec{k}_0}(\vec{q})$ has to vanish at least for one $\vec{q}$ in the Brillouin zone. Since, otherwise, we could use $\ket{u^*_+(-\vec{k}_0)}$ as a (momentum $\vec{q}$ independent) ``trial Bloch state'' that, upon projection to the band $\{u_+(\vec{k}_0+\vec{q})\}_{\vec{q}}$, would define a smooth and periodic gauge for that band. This would be at odds with the non-zero Chern number \cite{PhysRevB.74.235111}. 

\begin{figure}
    \centering
    \includegraphics[width=\linewidth]{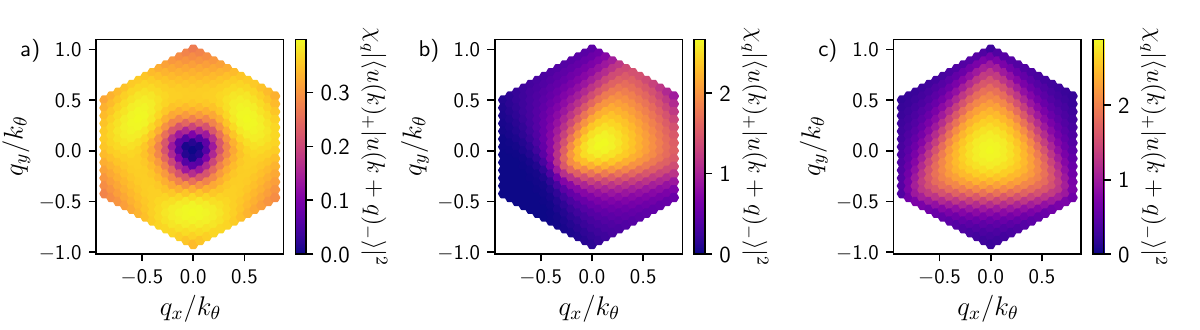}
    \caption{We show the quantity $\chi_{\vec{q}}|\langle u_+(\vec{k})|u_-(\vec{k}+\vec{q})\rangle|^2$ as a function of $\vec{q}$ in the first Brillouin zone for three different values of $\vec{k}$. In (a) we choose $\vec{k}=\Gamma$, in (b) $\vec{k}$ to lies halfway between $\Gamma$ and $K$, and in (c) $\vec{k}=K$. We use $\chi_{\vec{q}}$ peaked at $\vec{q}=0$ as in the main text and $D/d=14.45$ mV/nm. }
    \label{fig:FFs}
\end{figure}

Three-fold rotational symmetry further implies that $s_{\vec{k}_0}(\vec{q}) = s_{C_{3z}\vec{k}_0}(C_{3z}\vec{q})$. Therefore, for high-symmetry choices of $\vec{k}_0$ with $C_{3z}\vec{k}_0 = \vec{k}_0$, these zeros are either pinned at high-symmetry $\vec{q}$, which are themselves $C_{3z}$ invariant, or appear in multiples of three. These expectations are confirmed by \figref{fig:FFs}, where the intervalley form factors for the active band at the Fermi surface are shown as a function of $\vec{q}$ at fixed $\vec{k}$. For the high-symmetry choices $\vec{k}=\Gamma$ and $K$, the zeros are indeed pinned at high-symmetry points, too. While generic $\vec{k}$ leads to zeros at generic positions. 

\begin{figure}
    \centering
    \includegraphics{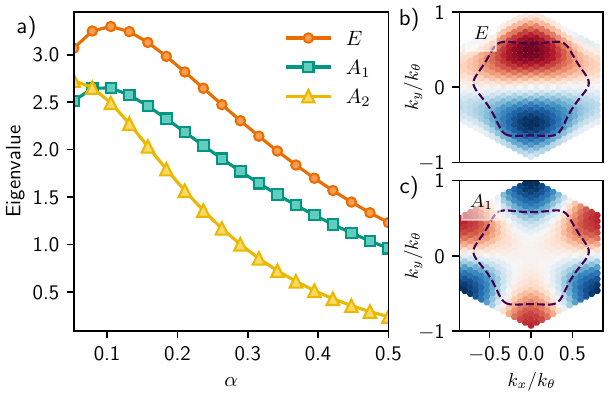}
    \caption{Solutions of the LGE for fluctuations of $\phi_{x,y}$ when we set the form factors $\lambda_{x,y}(\vec{k},\vec{q})=1$ to study the impact of the strong momentum dependence of $\lambda_{x,y}$ demonstrated in Fig.~\ref{fig:FFs}. As in \figref{fig:odd}, (a) shows the leading eigenvalues for each IR, while (b) and (c) show the corresponding order parameters in the Brillouin zone for the leading and subleading solution, respectively.  Comparison with \figref{fig:SCConnectedToStrong}(b) of the main text shows that removing the momentum dependence of the form factors favors the $E$ state relative to the $A_1$ state.}
    \label{fig:LGEnoFF}
\end{figure}

Based on the LGE in \equref{IVCGapEq}, it is natural that these topology-related momentum dependencies in the form factors influence the superconducting energetics. From \figref{fig:SCConnectedToStrong}(b), we can see that making the peak of $\chi_{\vec{q}}$ more concentrated around $\vec{q}=0$ by decreasing $\alpha$ favors the $E$ state. As such, it seems natural that the zero at $\vec{q}=0$ in \figref{fig:FFs}(a) is an important driving force for the $A_1$ state's dominance in \figref{fig:SCConnectedToStrong}. To test this hypothesis, we have removed the momentum-dependence of the form factors, $\lambda_{x,y} \rightarrow 1$, in the LGE and, indeed, this favors the $E$ pairing instability, see \figref{fig:LGEnoFF}. 


\end{appendix}
\end{document}